\documentclass[useAMS,usegraphicx]{mn2e}

\title[Star-forming Galactic Contrails]{Star-forming Galactic Contrails at z=3.2 as a Source of Metal Enrichment and 
Ionizing Radiation\thanks{
The data presented herein were obtained at the W.M. Keck Observatory, which is operated as a scientific partnership among the California Institute of Technology, the University of California and the National Aeronautics and Space Administration. The Observatory was made possible by the generous financial support of the W.M. Keck Foundation.}}
\author[Michael Rauch et al.]{Michael Rauch$^{1}$, George D. Becker$^{2}$, Martin G. Haehnelt,$^{2}$, Jean-Rene Gauthier$^{3}$ \\
$^{1}$Carnegie Observatories, 813 Santa Barbara Street, Pasadena, CA 91101, USA\\
$^{2}$Institute of Astronomy and Kavli Institute for Cosmology, Cambridge University, Madingley Road,  Cambridge CB30HA, UK\\
$^{3}$California Institute of Technology, Pasadena, CA 91125, USA}

\begin{document}


\pagerange{\pageref{firstpage}--\pageref{lastpage}} \pubyear{2011}

\maketitle


\label{firstpage}

\begin{abstract}  
A spectroscopically detected Lyman $\alpha$ emitting halo at redshift 3.216 in the GOODS-N field 
is found to reside at the convergence of several  Ly$\alpha$ filaments. HST images show that
some of the filaments are inhabited by galaxies. Several of the galaxies in the field have  pronounced head-tail structures,  which 
are partly aligned with each other. The blue colors of most tails suggest the presence of young stars, with the emission from at least one of the galaxies apparently dominated by
high equivalent width Ly$\alpha$.  Faint,  more
diffuse, and similarly elongated, apparently stellar features, can be seen over an area with a linear extent of at least 90 kpc.
The region within several arcseconds of the brightest galaxy exhibits spatially extended emission by HeII,  NV and various lower ionization metal lines.  The 
gas-dynamical features present are strongly reminiscent of  ram-pressure stripped galaxies, including 
evidence for recent star formation in the
stripped contrails. Spatial gradients in the appearance of several galaxies may represent a stream of galaxies passing from a colder to a hotter intergalactic medium.  The stripping of gas from the in-falling galaxies, in conjunction with the occurrence of star formation and stellar feedback
in the galactic contrails  suggests a  mechanism for the metal enrichment of the high redshift intergalactic medium that does not depend on long-range galactic winds, at the same time opening  a path for the escape of ionizing radiation from galaxies.

 \end{abstract}


\section[]{Introduction}

Long slit, spectroscopic blind surveys targeting the HI Ly$\alpha$ emission line have the potential to deliver detailed and otherwise unavailable insights into the gas dynamics
and, in conjunction with deep, space-based imaging, the star-gas interactions in proto-galactic halos and the intergalactic medium. Several surveys of this kind 
(Rauch et al 2008, Rauch et al 2011, paper I; Rauch et al 2013a, paper II; and Rauch et al 2013b, paper III) have discovered  a distinct subpopulation of
of extended, asymmetric Ly$\alpha$ emitters at $z\sim 3$,  with a comoving space density on the order of $10^{-3}$ Mpc$^{-3}$ and typical observed line fluxes of a few times $10^{-17}$erg cm$^{-2}$ s$^{-1}$. With  a large number of processes capable of producing Ly$\alpha$ radiation, one may expect the emitters to be drawn from a highly
inhomogeneous group of objects. However, the selection by Ly$\alpha$ emission is likely to favor galaxies in certain phases of their formation, when the stellar populations and gas dynamics are particularly conducive to the production and escape
of Ly$\alpha$ photons.  The peculiar spatial distribution and clustering behaviour of Ly$\alpha$ emitters  suggests that  
environmental effects and interactions may play an important role in determining whether a galaxy appears as a Ly$\alpha$ emitter (e.g., Hamana et al 2004; Hayashino et al 2004;  Kovac et al 2007; Cooke et al 2010, 2013; Zheng et al 2011; Matsuda et al 2012). Indeed, all four extended emitters  described so far in the papers in this series exhibit signs of interactions. 

The duration of the processes  leading to the production of ionizing radiation (e.g., the lifetimes  of massive stars, AGN activity), as well as   the astrophysical timescales relevant for the emission of Ly$\alpha$   in high redshift gaseous halos
(recombination- and resonance-line radiative-transfer time scales) tend to be short compared to  the dynamical timescales and life times of the general stellar population.
Thus  the spectroscopic detection of such a halo amounts to a "snapshot" of a particularly interactive phase in their formation, illuminated  by
a "flash" of Ly$\alpha$ emission.

Among those extended Ly$\alpha$ emitters published to date,
the first one showed diffuse stellar features, in addition to
a clear detection of the infall of cold  gas into an ordinary  high redshift galaxy (paper I).
Paper II described what may be a Milky-way-sized halo with multiple galaxies hosting disturbed, partly young stellar populations.  A thin filament apparently dominated
by high equivalent width Ly$\alpha$ emission  may reflect recent intra-halo star formation in a tidal tail or in the wake of a satellite galaxy. The third object, revealing the only case in this sample clearly  related to non-stellar processes, is an AGN illuminating a satellite galaxy,  possibly triggering the formation of very young stars in its halo (paper III). 
The object examined in the present paper is a large halo surrounded by Ly$\alpha$ filaments illuminated by a group of distorted, mostly blue galaxies.
As we shall argue below, the interaction in this case appears to be between the galaxies and a gaseous medium through which they move, and which appears to strip off part of their gas and induce  star formation in their wake.  

The observations are described in the next section, followed by a description of the galaxies coinciding  with the gaseous halo.
The presence of spatially extended metal emission, the nature of the Ly$\alpha$ filaments and the energetics of the emission are then discussed. The paper concludes with a discussion of the likely nature of the phenomenon and its significance for the metal enrichment  of the intergalactic medium and for the escape of photons responsible for its ionization. 
 
\begin{figure}
\includegraphics[scale=.4,angle=0,keepaspectratio = true]{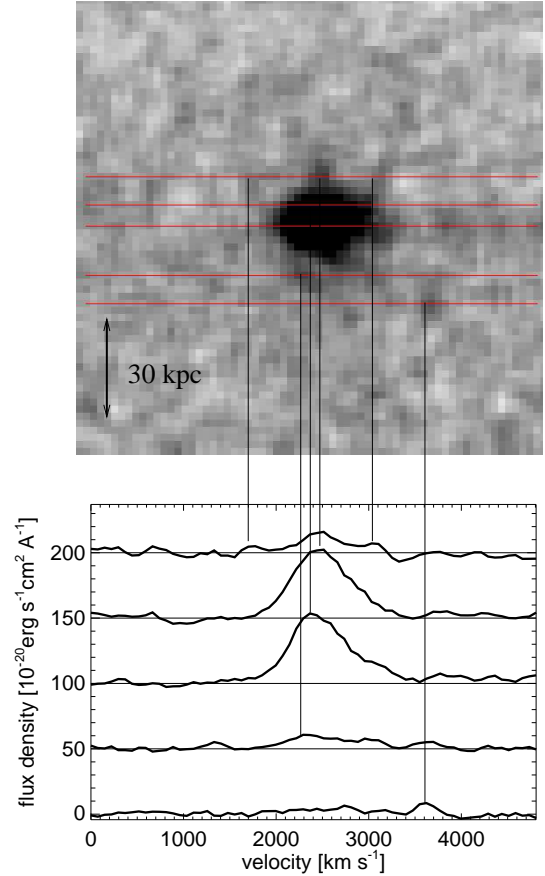}
\caption{Two-dimensional spectrum of the Ly$\alpha$ emission line (top panel). The spatial extent along the slit (vertical direction) is 17.3".
The bottom panel shows one-dimensional spectra along the five cuts delineated in the top panel,
extracted over 3 spatial pixels (0.81").
\label{cuts}}
\end{figure}

\begin{figure}
\includegraphics[scale=.5,angle=0,keepaspectratio = true]{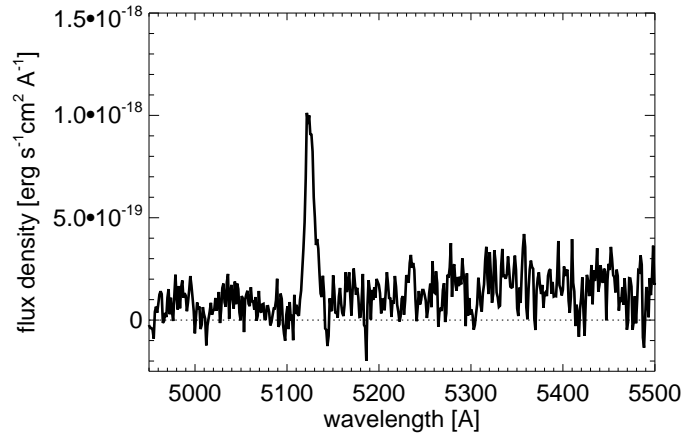}
\caption{One-dimensional optimally extracted spectrum.\label{onspec}}
\end{figure}

\section[]{Observations}


\begin{figure*}
\includegraphics[scale=.5,angle=0,keepaspectratio = true]{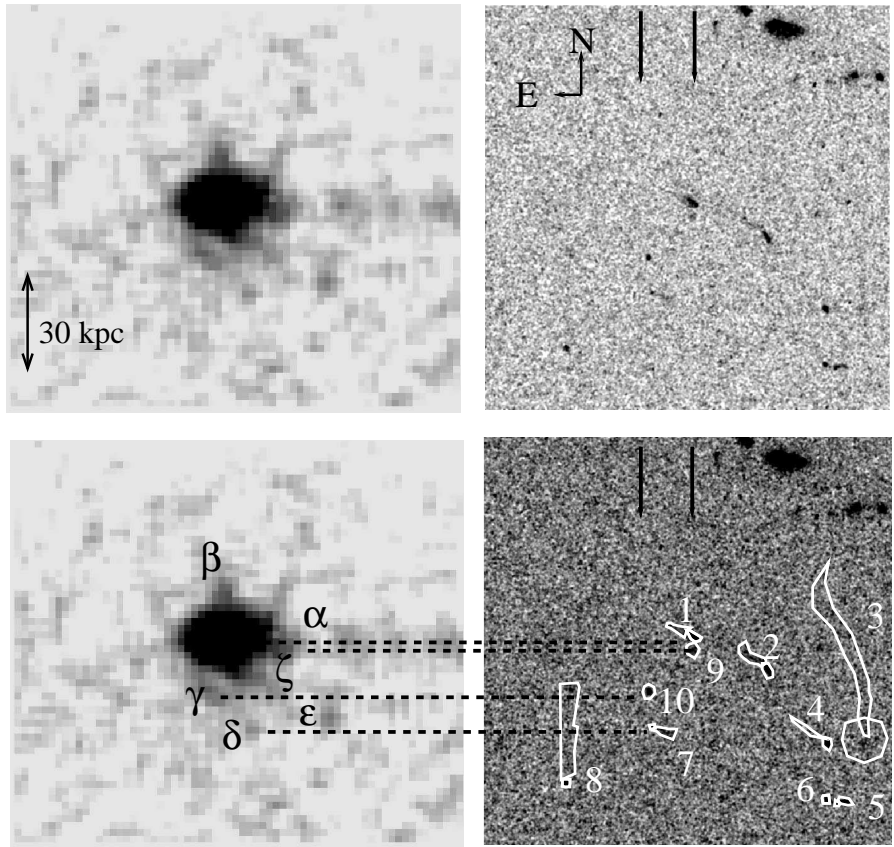}
\caption{Two-dimensional Keck LRIS spectrum of the Ly$\alpha$ emission line region (left two panels) and ACS B-band (F435W) image (right two panels; smoothed with a 3x3 pixel boxcar filter). In the spectrum, the dispersion
runs from left (blue) to right (red), and the N-S direction from top to bottom, with a spatial extent of 16".
For clarity, the bottom row of panels repeats the top row, but with annotations. The numbered white closed lines are the extraction windows used to obtained the flux measurements
given in table 1. The numbers refer to galaxies with broad band detections listed in table 1, the greek letters to features
in the Ly$\alpha$ spectrum. The two black vertical lines at the top of the images indicate the approximate position of the slit, which was determined
from a scheme minimizing the variance between the image collapsed perpendicular to the slit position, and the spectrum. The spectrum on the left 
is shown over a velocity stretch
of 4880 km s$^{-1}$ and shows apparent Ly$\alpha$ filaments extending south from the continuum position out to at least 27 kpc   proper  (from the center of the emitter to position $\delta$) in projection along the slit,
and further, faint protuberances appear to be sticking out to the north.   
The broad band smudges seen in the image  extend over at least about 90 kpc in the N-S direction and 
E-W directions. Several of the emission patches in the spectrum have counterparts in the B-band image (black dashed lines). \label{specplusim}}
\end{figure*}

\begin{figure*}
\includegraphics[scale=.6,angle=0,keepaspectratio = true]{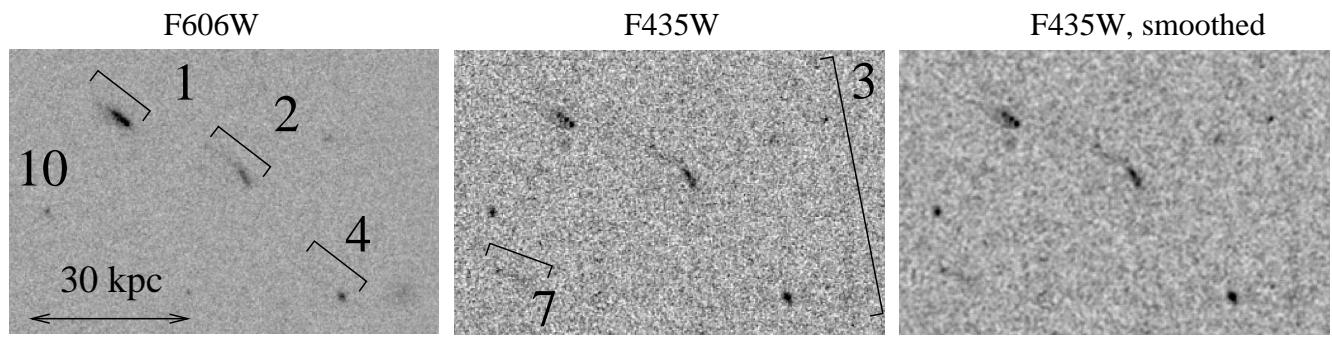}
\caption{Broad band images of several distressed galaxies to the SW of the position of the Ly$\alpha$ halo. The size of each image is $10.2" \times 6.7$", with N up and E to the left. The image
is centered at 12:36:46.48 +62:16:06.16 (2000).
The panels from left to right are: the ACS F606W (V band) image, the F435W (B band) image, and a version of the latter smoothed with a 3x3 boxcar filter.
Object 1, less than an arcsec to the W from the center of the slit,  is the source of most of the 
continuum emission and possibly most of the Ly$\alpha$ in the spectrum. 
Brackets show the multiple apparent head and tail structures for several objects.  For objects 1, 2, 3 and 4 the head appears to the SW of the tail. Object 10 is a point source, whereas 7 has a less distinct shape,
with a more E-W orientation and the peak flux occurring to the E. Note the multiple bright cores in the two tadpoles 1 and 2, seen in the rightmost panel. The V-band flux of object 4 may be dominated by Ly$\alpha$ emission. \label{tads}}
\end{figure*}

\begin{table*}
\scriptsize
 \centering
 \begin{minipage}{170mm}
  \caption{Properties of the sources}
  \begin{tabular}{@{}rcccccccc}
\hline 
 ID  & GOODS-N$^a$  & V$^a$ & B-V$^a$ & V (Head)  & B-V (Head) & V (Tail) & B-V (Tail)& B-band SFB (Tail) [mag/$\sq"$] \\
 \hline
1 & J123646.84+621608.1  &  24.90$\pm$0.03 & 1.10$\pm$0.11  &    25.60$\pm$0.06  &    1.27$\pm$0.10   &  27.07$\pm$0.09 &  0.79 +/-0.19    & 26.32\\
2 & J123646.42+621606.8  &  26.08$\pm$0.09 & 0.10$\pm$0.15  &    26.43$\pm$0.07  &    0.42$\pm$0.11   &  27.04$\pm$0.09 &  0.16 +/-0.16    & 26.41\\
3 & J123645.87+621604.0  &  26.27$\pm$0.09 & 0.88$\pm$0.27  &    26.45$\pm$0.10  &    0.56$\pm$0.25   &  27.14$\pm$0.24 & -0.37 +/-0.32    & 28.12\\
4 & J123646.08+621603.9  &  26.82$\pm$0.10 &  0.12$\pm$0.18 &    27.20$\pm$0.15  &   -0.14$\pm$0.20   &  28.20$\pm$0.20 &  $>$2.80$^b$        &  -  \\
5 & J123645.99+621601.6  &  27.54$\pm$0.14 &  0.09$\pm$0.24 &    28.18$\pm$0.15  &    0.07$\pm$0.21   &  27.89$\pm$0.13    & -0.01$\pm$0.19 & 25.75 \\
6 & not det.        & -               &             -  &    27.64$\pm$0.13  &   -0.01$\pm$0.17   & point source & -    & 27.82\\
7 &   not det.           &      -          &       -        &         -          &          -         &  31.12$\pm$2.68 &  -3.53$\pm$2.68  & 26.22 \\
8 &   not det.      &      -          &       -        &    27.15$\pm$0.11  &    0.53$\pm$0.16   &      $>$ 29.01     &    $<$ -0.69$^c$  & 28.04     \\
9 & not det.          &                 &                &    28.95$\pm$0.90 &   -0.89$\pm$1.00    &  point source &  - &  28.57\\ 
10 & J123647.09+621605.9  &  27.81$\pm$0.18 & -0.08$\pm$0.27 &    27.92$\pm$0.18  &   -0.39$\pm$0.24   &  point source      &      -        &  -  \\
\hline
\end{tabular}
\\
comments: a) Giavalisco et al 2004; b) adopting $1\sigma$ flux error for B; c) adopting $1\sigma$ flux error for V\end{minipage}
\end{table*}

The observations consist of a long slit, spectroscopic blind survey, with the slit positioned in precise N-S orientation on the object J123647.05+621237.2 in the Hubble Deep Field North (HDFN). Data were taken with the LRIS (Oke et al 1995; McCarthy et al 1998, Steidel et al 2004) 
B and R arms and the D560 dichroic,
using the 600/4000 grism in 2x2 binning (blue side) and the 600/7500 grating in 1x1 binning (red side), through a  custom long slit built from two slit segments with a combined size of 2" $\times$ 430".
Total exposure times of 35.8 hours in the blue and 35 hours in the red arm of LRIS were obtained in March 2008, May 2008, and April 2009. The resulting 1-$\sigma$ surface brightness
detection limit, measured for a $1-\sq"$ wide aperture, is approximately $1.1\times 10^{-19}$ erg cm$^{-2}$ s$^{-1}\sq"$.

The spectrum shows a strong Ly$\alpha$ emission complex near 5125.0 \AA\ (fig. \ref{cuts}), with the peak of the line corresponding to a redshift 3.2158.
The identification as Ly$\alpha$ is supported by the large spatial extent of the emission, and the drop in the continuum blueward of the emission
line, caused by the  Ly$\alpha$ forest (fig.\ref{onspec}). The Ly$\alpha$ line is broad (FWHM $\sim 910 $ kms$^{-1}$) and shows a red shoulder, with part of the width coming from superposition of multiple
sources (see below). After subtraction of the continuum trace, the total Ly$\alpha$ flux that passed through the slit is ($2.26\pm0.13)\times10^{-17}$ erg cm$^{-2}$ s$^{-1}$. 
Inspection of archival GOODS-N v2.0 images (Giavalisco et al 2004; fig. \ref{specplusim}, top right panel) shows 
the continuum source closest to the inferred slit position to be GOODS J123646.84+621608.1. A photometric
redshift of 3.09 (Xue et al 2011) for that galaxy, referred to as object 1 in the table and the figures, 
agrees well with our spectroscopic redshift. There is no prior redshift information for any of the other
faint galaxies, except for a few foreground objects in the N-W corner of the field. 

The Ly$\alpha$ emission line is remarkable in that several  filaments extrude to distances of at least 27 proper
kpc (in projection) away from the continuum position. At least one of these structures corresponds  to multiple features detected in the HST ACS F435W ("B band") (RHS panels of fig. \ref{specplusim}).
In an area extending to about 5 arcsecs on either side of the slit,  two highly distorted, tadpole-shaped galaxies (1 and 2; with object 1 producing the continuum in the spectrum) can be seen,
Both have substructure with multiple cores of blue light in their heads (figs.\ref{tads}, \ref{bowshock}).
Several more extended low surface brightness features can be seen further afield
(3,4,5,6,7) partly also with (less distinctive) head and tail structures. Among these are blue sources with (objects 5 and 8) and without (object 10) tails,
and at least one object (number 3)  with an apparent linear size of more than 40 kpc.  

Of the features in the Ly$\alpha$ line, the central emission  "$\alpha$" corresponds to tadpole 1 in the image,  "$\zeta$" to the smudge near number 9,  "$\delta$" to 7, and "$\gamma$" to  10. The latter two also have very faint continua recognizable in some regions of the spectrum
that line up very well with the Ly$\alpha$ emission, showing that the $\gamma -\delta$ filament is subtended by at least two galaxies.
The other significant Ly$\alpha$ smudges with greek letters, $\beta$ and $\epsilon$, plus the vaguely visible elongated  features on either side of $\beta$  may also have counterparts in the broad band, with $\epsilon$ possibly related to object 4 (see below).
The various white contours in the bottom right panel of fig. \ref{specplusim}, are marking extraction windows used to determine the fluxes listed in table 1. 
Even though the low surface brightness features are statistically significant detections (objects 3, 8,  and 7 are 5.0, 4.4, and 8.1 $\sigma$ excursions in the B-band) some of them could be observational artifacts. Moreover, even if they are real objects, they may not all be at the same redshift as the Ly$\alpha$. However, the spectroscopic identifications of objects 1, 7, and 10, and the simultaneous presence
of several highly distorted and/or  blue ($B-V\sim 0$) objects suggest that most of the features are likely to be real and related to the Ly$\alpha$ halo. Further distorted objects in the B band
occur to the N-W of the present region (not shown). If these belong to the same large scale structure (which, due to a lack of redshifts is uncertain), the total maximum extent could be as large as 210 kpc proper.
With the field being located near the edge of the GOODS-N imaging coverage we cannot follow the structure further to the North.

\section[]{Galaxy properties}

The various objects shown in the broad band figure are listed with their identifying numbers in column 1 of table 1, together with their name in the GOODS-N catalog (where detected) in
column 2.
The GOODS-N broad band B and B-V colors (columns 3 and 4) are followed in columns 5-8 by the same quantities determined directly in extraction windows placed on the head and tail
regions of the various objects shown in fig. \ref{specplusim}. Note that no corrections for extinction or intergalactic absorption have been applied.  From the work by Meiksin (2006), we expect that a
color correction of about  $\delta (B-V) \sim -0.4$ (for young star forming galaxies at z=3.2) needs to be added to the magnitudes in the table. 

The magnitudes of the "heads" show
good correlation with the GOODS-N "magbest" magnitudes where available for the same object. Because of the irregular apertures used, the absolute magnitudes are not precisely comparable among objects.  The useful information mainly lies in the B-V colors and their gradients between heads and tails.
Objects 2, 4, 5, 6, 7, 9, and 10 are generally blue (overall B-V$<$0.12 in the GOODS-N catalog, or, where uncatalogued, in our apertures). We caution not to read too much into the formally very blue colors as the sometimes
very large formal errors speak for themselves. However, there is a clear trend in that all the cases where there is a distinct head-tail shape (objects 1, 2, 3, 5, and 8), with the exception of object 4, have bluer tails  than heads.  The tendency for the tails to be bluer than the heads is different from the prevailing pattern for tadpole galaxies
(e.g., Elmegreen \& Elmegreen 2010).

\begin{figure}
\includegraphics[scale=.5,angle=0,keepaspectratio = true]{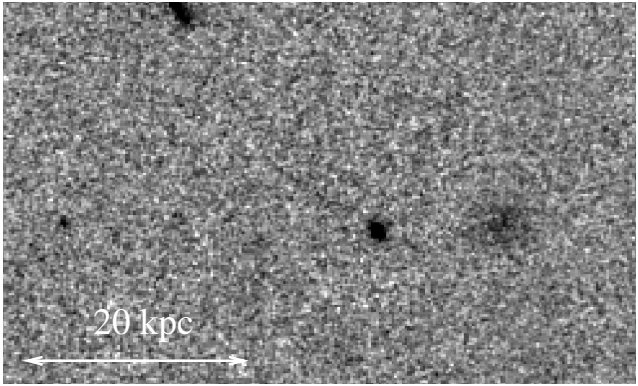}
\caption{$4.9\times7.3$ " V band image of object 4. \label{tad4}}
\end{figure}

The object 4 is an interesting case in that the tail is only visible in the V band (fig. \ref{tad4}), with a total flux of $(5.8\pm1.2)\times 10^{-17}$erg cm$^{-2}$ s$^{-1}$, corresponding to a 4.8 $\sigma$ detection. The situation here may be similar to the one presented in paper II, a narrow filament
of emission visible in only one broad band, that may be glowing in Ly$\alpha$, except that in the current situation, the Ly$\alpha$ line, due to its higher redshift, would be situated in the V band and not in B. Because of the low flux,
other broad band images do not provide useful constraints  (e.g., V-I = $-1.3\pm 1.0$). We can, however, put limits on the equivalent width from the B and V band fluxes, assuming that the B band measures the continuum,
and the V band the continuum plus line flux, with the same assumptions as used in paper II. As the equivalent width is directly dependent on the ratio of V-band flux to B band flux,
we estimate an equivalent width lower limit by setting the V-band flux to its $-n\sigma$ excursion, and the B-band flux (which is formally measured to be zero) to its positive, $+n\sigma$ value. The resulting rest frame equivalent width lower limit 
for n=1 is $EW_r > 1130 \AA $. For n=2, the value is already a rather moderate 57\AA , and the "3$\sigma$" result is compatible with both the V and B band just containing pure continuum emission of a flat spectrum source.
However, if the observed V band flux were mostly Ly$\alpha$ emission, the $(5.8\pm1.2)\times 10^{-17}$erg cm$^{-2}$ s$^{-1}$ should be easily seen in a spectrum as it is more than two times the flux of in the main halo.
In this case, the fact that the western edge of the slit is about 3.9" away from the tail of 4  is likely preventing us from detecting a strong Ly$\alpha$ emission line. It is intriguing that the Ly$\alpha$ feature $\epsilon$, though stronger than
either $\delta$ or $\gamma$, which both correspond to galaxies clearly detected, does not have a corresponding broad band object near enough to the slit to explain the relatively strong signal. Thus $\epsilon$ may just be 
a small fraction of the Ly$\alpha$ emission from the tail of  object 4, spilling over into the slit.

\begin{figure}
\includegraphics[scale=.75,angle=0,keepaspectratio = true]{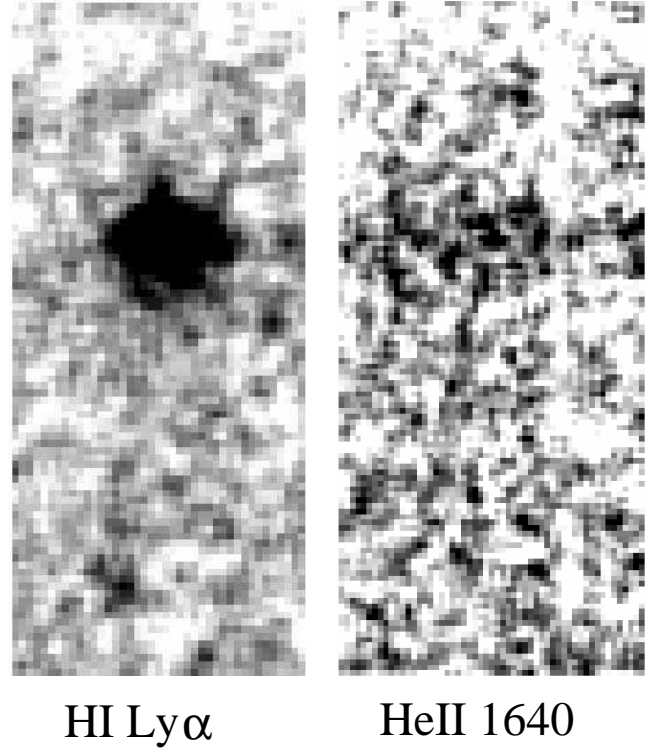}
\caption{Part of the spectrum showing again the HI Ly$\alpha$ line region, and the corresponding region from the LRIS
red arm spectrum, containing HeII 1640. The sections have been scaled to have the same velocity (horizontal; width 2960 km s$^{-1}$) and spatial extent (vertical; 25.6"). The spatial center of the sections is
somewhat offset to the south of the continuum trace in order to include an apparent spur of emission ending in a bright spot about 13" 
to the south. The HeII emission is somewhat effected by sky subtraction residuals.
\label{he_hi_portraits}}
\end{figure}

\subsection[]{Spatially extended line emission} 

\subsubsection[]{HeII 1640}

Aside from the Ly$\alpha$ emission, several other extended emission features can be seen in the spectrum.
He II 1640.4 \AA\ is noisy but clearly present near an observed wavelength of 6914 \AA. In the spatial direction it ranges  over several arcsecs (fig. \ref{he_hi_portraits}). The presence of multiple residuals from the subtraction of sky lines  and a strong spatial gradient in the sky background
make a flux measurement difficult, but we estimate that the flux is about $(4.5\pm1.4)\times10^{-18}$ erg cm$^{-2}$s$^{-1}$ (statistical
errors only), within a distance of 4.3" on either side of the continuum trace. Formally, this is approximately 19\% of the HI Ly$\alpha$ flux.
This comparison with HI should not be taken too serious as
HeII 1640 is likely to be less optically thick than HI Ly$\alpha$, so its spatial width  may largely reflect the spatial extent of the emitting gas, whereas a substantial fraction of HI Ly$\alpha$ may have been scattered further outside of the slit. Thus, the observed HeII 1640/ HI Ly$\alpha$ flux ratio may be a considerable overestimate. 
If we approximate the spatial profile of the Ly$\alpha$ halo by a circular Gaussian profile with the radius set to the spatial extent {\em along} the slit (FWHM=1.9"),  only about 26\% of the Ly$\alpha$ emission would have passed
through the slit. This could reduce the HeII/Ly$\alpha$ to 0.05, a value  consistent with several different ionization mechanisms (see the discussions in Yang et al 2006; Prescott, Dey \& Jannuzi 2009,
and Scarlata et al 2009).

\begin{figure*}
\includegraphics[scale=.5,angle=0,keepaspectratio = true]{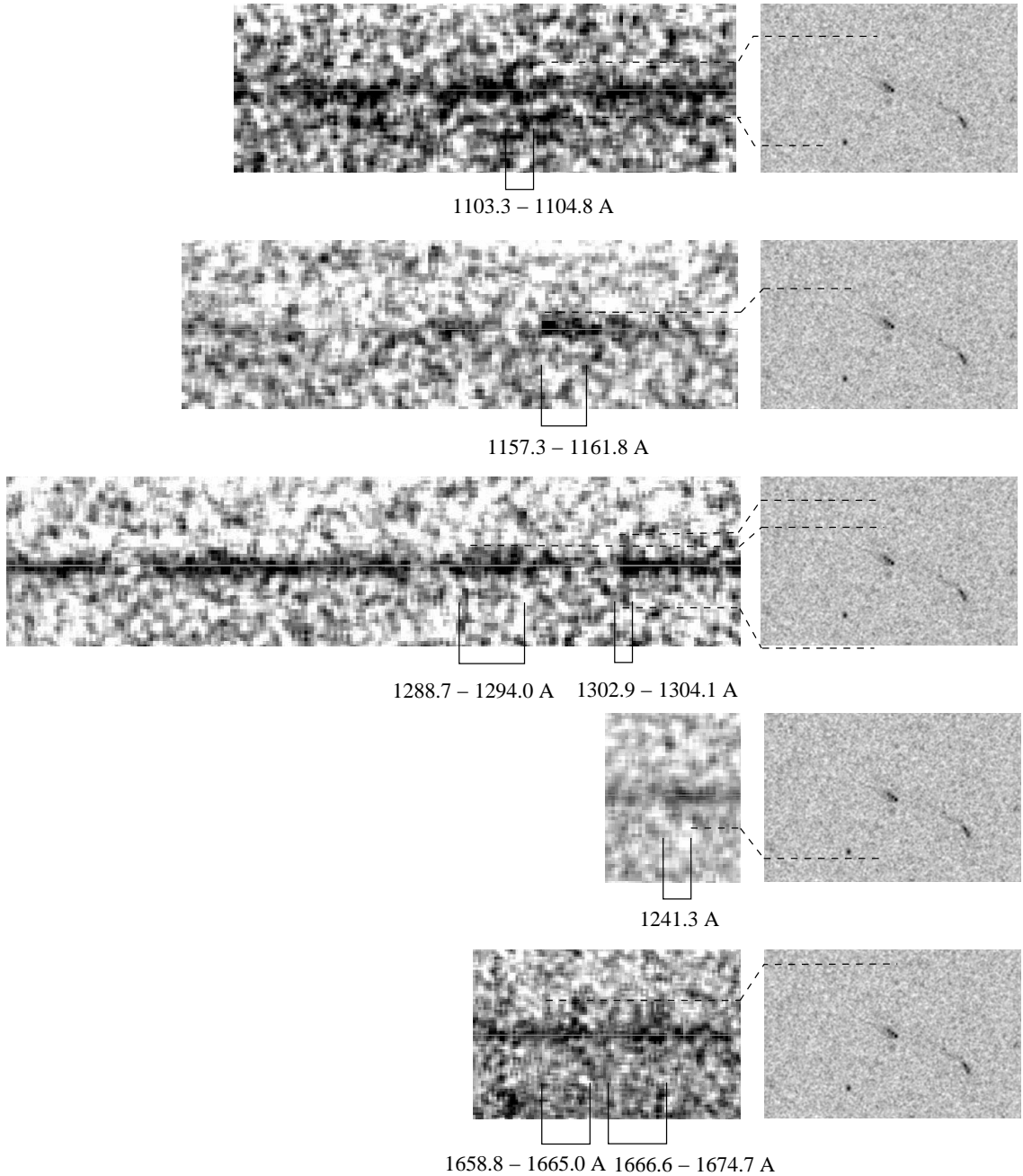}
\caption{Several individual stretches of the spectrum, showing spatially extended emission features with their rest wavelengths (assuming z=3.2158), and 6.7" high ACS F435W images centered
on tadpole 1.  The horizontal, thin, light gray solid lines indicate the median position of the continuum in the spatial direction.
The black, dashed lines are meant to guide the eye as to the extent of the spectral emission in relation to  tadpole 1. The region 1302.9-1304.1 \AA\ is a sharp wavelength feature that extends out on both sides of the continuum, whereas other features are more asymmetric. The significance and lateral extent of some of the regions can be judged from
fig. \ref{spatial_cuts}.  
\label{juxtaspec}}
\end{figure*}

\subsubsection[]{Metal transitions and extended emission}

Several features belonging to metal line transitions can be seen in the one-dimensional spectrum (fig. \ref{onspec}), including an absorption/emission complex, with the peak emission occuring near near 5234.6 
\AA\ (FWHM $\sim 500$ kms$^{-1}$), probably related to the NV 1238/1242 \AA\ doublet, and absorption troughs at 5302.5 \AA\ and 5418.8 \AA.  We identify the lower redshift
one with the SiII 1260 \AA\  resonance line.
A further, broader trough near 5480 \AA\ almost certainly corresponds to the OI/SiII 1302-1304 \AA\ complex often seen in Lyman break galaxies (e.g., Shapley et al 2003). The low
signal-to-noise ratio does not permit a very meaningful analysis, but  single component, Gaussian fits to the absorption troughs at 5302.5 \AA\ and 5418.8 \AA\ show them to be blue-shifted with respect to the nominal redshift from the peak of the Ly$\alpha$ line, z=3.2158, by about
-630 and -500 km s$^{-1}$, with rest frame equivalent widths on the order of 3 \AA. 

\begin{figure}
\includegraphics[scale=.49,angle=0,keepaspectratio = true]{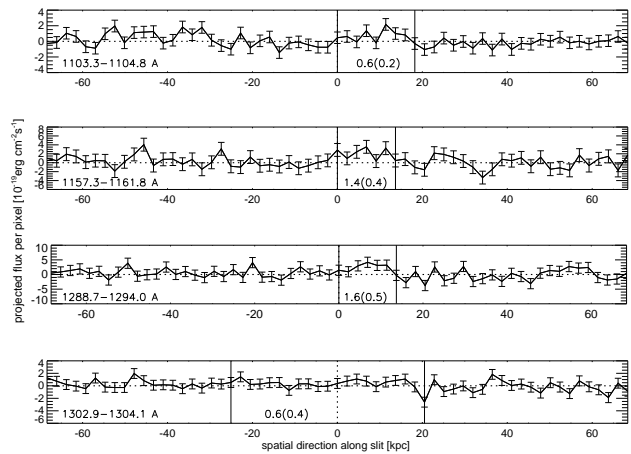}
\caption{This figure gives an idea of the fluxes, spatial extent and asymmetry of several of the extended emission line features shown in the previous
figure. It shows
spatial profiles of the first four spectral segments from fig.\ref{juxtaspec}, with the origin placed at the median position of the continuum profile. The regions of the spectral features shown in fig. \ref{juxtaspec} were collapsed along the dispersion direction,
and a mean continuum obtained from a nearby wavelength region was subtracted. 
The two vertical solid lines in each profile show roughly where the edges of the residual visible emission occur in the 2-d spectrum.
The total fluxes as measured between the positions along the slit indicated by the vertical lines are given (together with
their $1\sigma$ statistical errors in brackets) in units of $10^{-19}$ erg cm$^{-2}$ s$^{-1}$.  
\label{spatial_cuts}}
\end{figure}

In several wavelength regions below and above Ly$\alpha$, the spectral trace seems to be broadened, suggesting spatially extended emission line regions. We show some of these
regions in (fig. \ref{juxtaspec}), indicating the spatial correspondence between spectrum and image  by dashed lines.  To give an idea of the flux levels  and 
significance of the excess emission along the slit,  fig. \ref{spatial_cuts} shows some of the spatial profiles after collapsing the regions in the previous figure along the dispersion direction. Note that an average continuum spectrum has been subtracted in fig.\ref{spatial_cuts}, and the fluxes  represent just the excess emission. 

\medskip

Because of the low signal-to-noise ratio, the presence of Ly$\alpha$ forest absorption, and the unknown mechanisms populating the atomic levels
we found it difficult to assign the extended emission to definite transitions. The few identifications suggested here are tentative and may change with
better data:

The first region (top panel in fig. \ref{juxtaspec}) comprises relatively prominent emission extended to both the north and south of tadpole 1.
The spectral traces of two other objects, 10 and 7 in fig.\ref{specplusim},  are  visible  as well just below the trace of the main galaxy for a wavelength stretch of about 50 \AA\ in the rest frame, of which we show the central part in this panel. 
Having enhanced emission at the same wavelength as the main tadpole strengthens our proposed identification of these objects with galaxies 
sharing the same redshift, responsible for the  Ly$\alpha$ emission in the filaments $\gamma$ and $\delta$. 
Among the transitions in this wavelength range that could produce the band-like spectral character, possible candidates are the numerous FeII and FeII] lines in the vicinity of 1104 \AA, with several transitions from the ground-state and accompanying low-lying excited states.

The second panel from the top, shows an asymmetric, extended emission region (1157.3 - 1161.8 \AA) redward of an absorption trough. 
The nature of this feature is unclear.

In the third panel of  fig.\ref{juxtaspec} we find deep absorption troughs just blueward of extended emission near 1288.7-1294.0 \AA\  and
1302.9-1304.1 \AA. The identification of the lower redshift system is uncertain, but
the spectrally narrow, spatially very extended emission region 1302.9-1304.1 \AA\   most likely corresponds to the OI 1302.17, 1304.86, 1306.02 \AA\ triplet and the SiII 1304.37, 1309.28 \AA\ doublet (with possible contributions by other ions) which would also account for the absorption trough immediately blueward. 
SiII may be expected to be the dominant contributor under a wider range of possible physical scenarios, but is perhaps somewhat too far to the red of the optimal
position. The spatial extent of the emission
appears to exceed that of the other emission regions discussed earlier in both directions along the slit. If the emission is due to OI, it could be enhanced by Bowen fluorescence (Bowen 1947), i.e., pumping of OI by HI Ly$\beta$. A situation like this could arise in the contact zone between partly neutral gas (OI)
embedded in more highly ionized gas where recombinations produce Ly series photons.

The fourth panel shows the NV 1238,1242 \AA\ doublet region.  NV  appears to be present in the form of an absorption trough, as commonly seen in high redshift galaxies, plus
NV emission (marked as "1241.3 \AA"), which is extended to the south. At a flux of $(1.8\pm1.2)\times10^{-18}$ erg cm$^{-2}$s$^{-1}$, a comparison with other metals commonly seen in ionized gas, is difficult. The flux at the CIV 1548,1549 \AA\ doublet position (not shown),  which is commonly stronger than NV,   at $(0.83\pm1.5)\times10^{-18}$
erg cm$^{-2}$s$^{-1}$ is not significant.

In the bottom panel, a wavelength region from the red side of the spectrum contains a comb-shaped, multiple emission line pattern extending about 2.7" (21 kpc proper) to the north,
unfortunately marred by two background residuals.
We tentatively identify the lines between 1658.8 - 1665.0 and 1666.6 - 1674.7 \AA\  with OIII] 1660, 1666 \AA\ (which most  likely would be collisionally excited) and AlII 1671 \AA\ (which could be pumped by continuum radiation). 

\section[]{Origin of the Ly$\alpha$ emission}

The uncertain spatial extent of the Ly$\alpha$ emission beyond the slit prevents a measurement of the escape fraction of Ly$\alpha$, but
we can establish whether there is a sufficient source of Ly$\alpha$ that can explain the observed flux.
If due to photoionization, the observed flux of Ly$\alpha$ photons requires an ionization rate  of about $1.98\times10^{53}$ erg cm$^{-2}$ s$^{-1}$Hz$^{-1}$. 
Under the assumptions made in paper I and II, this can be achieved by a star-forming galaxy with rest frame luminosity $L_{1500\AA}$ = $2.34\times10^{28}$ erg s$^{-1}$Hz$^{-1}$
or V band (AB) magnitude = 26.4. With the brightest galaxy, number 1, having a  magnitude of 24.9, there nominally are about four times as many
ionizing photons available as required to explain the observed Ly$\alpha$, not counting any of the other sources in the field. 
Even if the slit losses approached a factor $\sim 4$, as discussed above, stellar photoionization would remain a viable explanation.
Based on the existing data we cannot rule out the presence of an AGN, but there currently is no positive evidence for such an object either.
 
The star formation rate associated with galaxy 1, if estimated by the usual relation (Madau, Pozetti \& Dickinson 1998), is found to be
\begin{eqnarray}
SFR=2.9 M_\odot {\rm yr}^{-1} \times\frac{L_{1500}}{2.3\times10^{28} {\rm erg\ s}^{-1}{\rm Hz}^{-1}}.
\end{eqnarray}

For some of the more clearly circumscribed emission regions in the individual filaments the photon budget  can be studied independently.
For example, the tail of region 7, with a V band magnitude 31.12, coincides in projection with the Ly$\alpha$ spot $\delta$ in one of the southern filaments.
The observed Ly$\alpha$ flux of that spot requires V=30.9, which is very close to the required value for a situation where virtually all ionizing radiation is being trapped
and converted into Ly$\alpha$ photons, with all photons escaping.

\medskip

The Ly$\alpha$ filamentary pattern is at least partly related to point sources that appear connected 
to the central emission peak by bridges of Ly$\alpha$ emitting gas. In principle, these bridges could consist of ionized gas, in which case the velocity position of the
corresponding Ly$\alpha$ emission should be close to the systemic velocity. We cannot test the ionization state of the gas for the individual filaments, but for the main
galaxy we have additional evidence from the absorption troughs present, which are all blue-shifted by on the order of $|v_{abs}|\sim 500-600$ kms$^{-1}$ with respect to the Ly$\alpha$ emission peak. 
In the case of Ly$\alpha$ emitters associated with Lyman break galaxies, the redshifted velocity of the Ly$\alpha$ emission line with respect to the systemic redshift of the galaxy, is about 2-3 times
the absolute value of the (blue-shifted) outflow velocity  $|v_{abs}|$. This puts the redshift of the Ly$\alpha$ emission relative to the systemic redshift at between 250 - 600 kms$^{-1}$,
which is well consistent with the range observed for Lyman break galaxies (e.g., Rakic et al 2011), and would mean that at least the central Ly$\alpha$ peak is highly optically thick.  
Taken together with
the presence of an absorption trough in the 1-d spectra (fig.\ref{onspec}) near 5100 \AA, the evidence suggests that the Ly$\alpha$ emission
is indeed the red peak of a double humped profile, with the blue peak strongly suppressed.

\medskip

\section[] {Interpretation}

As for the nature of the galaxies in the field, most of the objects with a head-tail structure have their heads to the south of the tails (1,2,3,4 and 5), making it less likely
that the features have been produced mainly by  tidal interactions. Rather, the evidence appears consistent with large scale ram-pressure stripping (Gunn \& Gott 1972; Nulsen 1982)
of gas, and recent star-formation in the down-stream ablated tails.

\begin{figure*}
\includegraphics[scale=.5,angle=0,keepaspectratio = true]{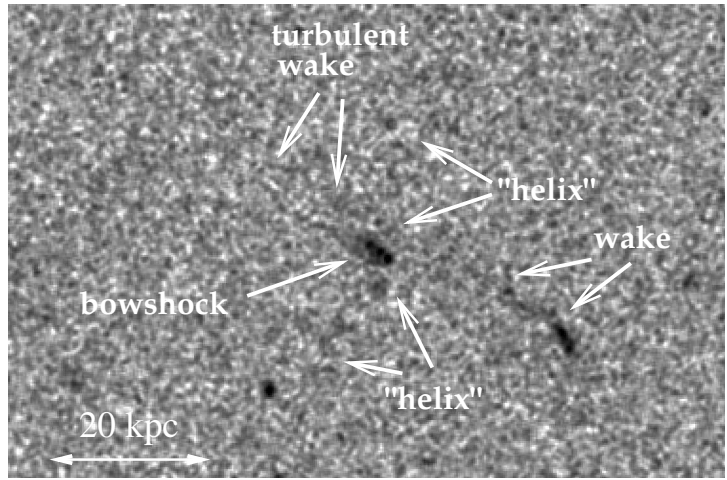}
\caption{Gasdynamical features indicative of interactions between the galaxies 1 and 2 and the intergalactic medium. The height of the image is 7.5".\label{bowshock}}
\end{figure*}

Ram pressure stripping has been invoked to explain features seen in several astrophysical environments, including galaxy clusters (e.g., Cayatte et al 1990; Vollmer \& Huchtmeier 2003; Chung et al 2007; Cortese et al 2007;  Hester et al 2010;  Smith et al 2010;  Yagi et al 2010;
Fumagalli et al 2011; Fossati et al 2012; Arrigoni-Battaia et al 2012), the Milky Way halo (e.g., Lin \& Faber 1983), the Local Group (e.g., McConnachie et al 2007) and  in galaxy groups (e.g., Rasmussen, Ponman \& Mulchaey 2006; Marcolini, Brighenti, \& D'Ercole 2003).  So far, there is relatively scant observational evidence for this process at high redshift, with the  exception of the 
tadpole galaxies, a population of galaxies increasingly common with redshift (e.g., Straughn et al 2006), that may partly have been shaped by ram-pressure (Elmegreen \& Elmegreen 2010).

Groups of galaxies falling into clusters seem to produce $H\alpha$ morphologies similar to the galaxy contrails observed here (e.g., Cortese et al 2006; Owen et al 2006; Sun et al 2007; Yoshida et al 2012).  In the present case, 
a number of observational details closely resemble hydrodynamic features predicted from simulations of supersonic motions in galaxy clusters
(e.g., Roediger, Br\"uggen, \& Hoeft 2006; Zavala et al 2012):
the main tadpole 1 shows a shape reminiscent of a bow shock, with a projected tangential angle of close to 45 degrees (indicating, at face value,  motion with a moderate Mach number $\sim$ 1.5), and what looks like a turbulent wake consisting of a  vortex street with a longitudinal extent of about 12 kpc and a maximum width of about 5 kpc (fig. \ref{bowshock}). A faint, linear structure vaguely resembling a "helix" extends almost perpendicular to the axis of symmetry (i.e., the head-tail direction)  to both sides of the galaxy, with a maximum traceable
extent of $\sim 20$ kpc to the south and somewhat less to the north. The nature of this feature is not clear. Some of the extended emission seen in fig. \ref{juxtaspec}
appears to arise from that  region, suggesting that it may be associated with both low and highly ionized gas. 
The pattern with the "helical" appearance could perhaps arise through hydrodynamic instabilities. A broadening tail of vortices (e.g., Roediger, Br\" uggen, \& Hoeft 2006, their fig. 4), may look similar, if the tadpole were viewed almost frontally, moving toward
the observer, and the "helical" structure were on the far side of the tadpole. 
Similar, curly structures also occur in the wakes of {\em interacting} galaxies undergoing stripping (e.g., Kapferer et al 2008, their fig. 9). 
Alternatively, the apparent spiral pattern may suggest
a passing, rotating small galaxy with an asymmetric outflow of gas that traces out a
helical pattern. If the structure were an outflow from the tadpole galaxy 1, e.g., the helical jet of an AGN,  a large outflow velocity 
faster than the relative velocity between galaxy and ambient gas could explain why the "helices" are not being swept back at a sharper angle.

\medskip
Tadpole 2 does not show a discernible bow shock, and has a narrower wake. The object 4 that we associated above with high equivalent width Ly$\alpha$ emission 
exhibits a linear,  thin tail (fig. \ref{tads}, \ref{tad4}),
not unlike the expected outcome of Bondi-Hoyle accretion (Sakelliou 2000).    

The proximity of the galaxies to each other, the presence of a bow shock in one but not in the others, and the various degrees of turbulence in their tails may imply that the flow of galaxies is encountering inhomogeneous conditions.
Specifically, the spatial sequence from tadpole 4 to 2 to 1 (fig. \ref{tads}) could be interpreted as indicating passage through a zone with a significant temperature gradient (perhaps an accretion shock) from a colder to a hotter gaseous medium: 
the tadpole 4, which is most advanced in the direction of the flow, has a linear, apparently non-turbulent tail, suggesting that it may be experiencing the higher viscosity and thermal
pressure of a hotter environment (e.g., Roediger \& Br\"uggen 2008). Tadpole 2 is more turbulent, but the tail is still narrow, and it does not have a bow shock either. 
Tadpole 1 has a bow shock, presumably because it is on the colder side of the interface between hot and cold gas, and its velocity relative to the intergalactic medium
exceeds the sound speed for the colder gas. This condition would easily be satisfied 
when passing through the general filamentary IGM with  even moderate velocity, as the typical temperature at $z\sim3$ is only  a few times $10^4 K$ (Rauch et al 1996), corresponding to
sound speeds of a few tens of kms$^{-1}$.  The tail of tadpole 1  is flaring up and shows a turbulent vortex pattern, consistent with the lower viscosity expected in a colder medium.

\medskip
 
A particular interesting consequence of ram-stripping is the formation of stars in the stripped gas, which has been the subject of recent observational  (e.g., Yoshida et al 2012; 
Smith et al 2010: Hester et al 2010; Fumagalli et al 2011;  Boissier et al 2012; Owers et al 2012) and theoretical study (e.g., Kapferer et al 2009, Tonnesen \& Bryan 2012; see also the Ly$\alpha$ emitting filament in paper II, which may have a similar origin). Several strands of evidence suggest that the current situation is indeed
an instance of star formation in galactic wakes: the tails of most of the objects have blue colors, indicative
of very young stars. The  finding of extended metal line emission far out from the galaxies suggests the presence of a stripped, or re-created, interstellar medium that is being excited by the newly forming stars. The high Ly$\alpha$ equivalent width suggested by the tail of galaxy 4 may be another sign of hot, young stars,
as discussed in paper II. 

The usual condition for ram-pressure stripping to take place is that the ram-pressure on the gas in a galaxy, experienced when passing through the intergalactic medium, needs to exceed the gravitational binding force per surface area,
$\rho v^2 \geq (\pi/2) GM(< R) \rho_{\rm gal}(R)/R$ (e.g., McCarthy et al 2008). Here $\rho$ is the ambient gas density, $v$ the relative velocity of galactic gas and ambient intergalactic medium, and $G$, $R$, $M(<R)$ and $\rho_{\rm gas}$ are the gravitational constant, the distance of a given gas volume element from the center of the galaxy, the total gravitating mass internal to that radius,
and the galactic gas density, respectively. 
It has often been assumed that ram pressure stripping is most relevant for low redshift, massive clusters.
However,
the
hierarchical nature of structure formation, leading to more compact gravitational potential wells, higher gas densities, and higher interaction rates at $z\sim3$ (when compared to the local universe)  suggests that one should expect miniature versions of the ram-pressure stripping seen in low redshift clusters  to  occur among satellites in individual high z galactic halos. With the higher density at high redshift favoring a higher pressure for a given velocity, the in-falling satellites themselves collapse from a denser background as well, so for the effect of ram pressure stripping one would have to look  to lower mass, dwarf galaxies, perhaps aided by processes that may lower the binding energy of the gas further.
The higher merger rate at high redshift may also work to increase the amount of ram-pressure stripped gas (e.g., Domainko et al 2006;  Kapferer et al 2008), as may
stellar or galactic outflows, as long as they can offset a significant part of the gravitational binding energy.  In addition, new cosmological
hydro-simulation techniques suggest more efficient stripping (e.g., Hess \& Springel 2012) and the presence of puffed up, high-angular momentum gas (e.g., Keres et al 2012), "ready to go".

Recently it has been argued (Benitez-Llambay et al 2012) that, in particular, dwarf galaxies do not even require the encounter with fully formed massive halos but
can lose gas to ram-pressure stripping in large-scale structure filaments at high redshift when entering terminal nodes like the (future) Local Group pancake. In this case,
our spectrograph slit may have intersected a filament of the cosmic web, lit up by the star-formation in the ablated contrails of a swarm of coherently moving galaxies.
To attain the high relative velocities required for stripping, the galaxies would have to move highly supersonically with respect to the gas they are plunging into. While we appear to be seeing one galaxy with a bow shock, it is not clear if the velocities of the other objects are high enough for this to work. However,
as argued above, the presence of an accretion shock in the terminal node with hot gas on one side may make this scenario consistent with the observations.
A variant of this picture may explain the stripping and the apparent gradient in the properties of the tadpoles  as a group of galaxies being "hosed down" when obliquely passing an accreting stream of gas.

\subsection[]{Metal enrichment and escape of ionizing radiation from star formation in stripped gas}

The existence of such extended structures at high redshift, the relatively large number density
of galaxies with tidal or ram-pressure related features (see also the disturbed halos described in paper I and II), and the presence of multiple sites of star-formation  in a common gaseous halo or large scale filament suggest that the stripping of gas
from galaxies in interactions could be an important contributor to the metal enrichment of the intergalactic medium, analogous to the lower redshift process leading to the enrichment of the gas in galaxy clusters. To explain the finding of metal enrichment in the IGM at large distances
from the nearest bright galaxy, galactic winds
from Lyman break galaxies have been invoked to drive  metal-enriched gas far into intergalactic  space 
(e.g., Pettini et al 2001; Steidel et a 2010). Among the persistent uncertainties with this  scenario is that the actually observed ranges of  galactic winds invariably fall short
of accounting for the  metals seen in QSO absorption systems at large distances from such galaxies. However, if, as we have argued above, ram pressure stripping of in-falling dwarf galaxies
and star formation in the stripped wake operate at high redshift, there may be less
need to invoke long-range winds from the central galaxy of a halo.
In this alternative picture,
the ram-pressure that led to the ablation of gas and subsequent star formation may also act to dispel newly formed metal enriched gas from the tails, aided by stellar winds and supernova explosions
that would find it much easier to escape from the weakly bound (and dark-matter free) star forming regions of galactic wakes. In any case, differential motion between the lost gas and the parent star forming galaxies
will distribute the gas spatially over time, and the assumption that this process mostly occurs in in-falling dwarf satellites implies that the gas is automatically reaching 
distances from any brightest halo galaxy as large as commonly observed (in metal absorption lines; e.g., Chen 2012). Tidal interactions between satellites may lead to a similar 
result, metals expelled into the gaseous halo of brighter galaxies, that came from the shredded interstellar medium of its satellites or from outflows in tidal dwarfs, and both processes 
may exist among the extended, asymmetric Ly$\alpha$ emitters in our study. There may be differences between the metallicities of the gas ejected from tidal star-forming regions,
and the gas lost by star forming regions in galactic contrails. Stars in the former arise from the relatively metal-rich ISM of the parent galaxy, as would stars in
gas ablated by ram-pressure or viscous stripping. Stars forming in turbulent wakes behind the galaxies may feed on the lower metallicity gas in the
halo or intergalactic medium as well, which may contribute to the signatures of hot, young stars described in paper II.

As argued earlier in paper II, extragalactic star formation in the wakes of stripped galaxies and in tidal tails
may facilitate the production and escape of ionizing photons and may have brought about the reionization of the universe at high redshift.
Star formation outside of the dense galactic HI cocoons  would lead to lines of sight with reduced optical depth for ionizing photons.
The presence of young, massive stars in the wakes, with the weak gravitational binding force enabling easy removal of neutral gas by even moderate amounts
of stellar winds or supernova outflows would all tend to enhance the escape of ionizing radiation. Recently, Bergvall et al (2013), examining selection
effects in the search for local galaxies leaking ionizing radiation, have come to similar conclusions as to the likely conditions required, including the importance
of stripping.

\section[]{Conclusions}

We have detected a Ly$\alpha$ emitting halo with several faint filaments stretching over  tens of kpc. The filaments correlate with star-forming regions in the form of  mostly
blue, faint galaxies, several of which have a distinct tadpole shape and blue, partly  turbulent tails, with one object showing what appears to be a bow shock. The GOODS-N ACS F435W  image reveals many such features criss-crossing an area
several times bigger than the visible extent of  the Ly$\alpha$ halo. The emission of the central halo and of the filaments is broadly consistent with being powered by stellar
photoionization. We detect spatially extended emission lines from gas surrounding the main tadpole, including  HeII 1640, NV 1240 and probably OIII], AlII, and FeII, 
suggesting an extended, extragalactic, interstellar medium with current star formation. 

The tadpole shapes, partial alignment, and the considerable numbers of unusual broad band objects make it unlikely  that the features observed are predominantly tidal in origin (i.e., caused  by individual two-body encounters).   Instead,
the galaxies may have experienced stripping of gas when moving relative to the intergalactic or intra-halo medium, with stars forming downstream in the galactic contrails. This process is observationally and theoretically well established in the local universe. Our observations have identified an occurrence of ram-pressure stripping at high redshift,
possibly involving dwarf galaxies interacting with the gas in more massive, individual galactic halos. The filamentary structure trailing behind a galaxy in the 
z=2.63 halo described in paper II may be another example of this effect. In the present case,
the properties of several tadpoles change along their general direction of motion, which may be consistent with these galaxies  passing into a hotter gaseous environment,
possibly the region behind an accretion shock.
Such a stripping scenario may play out on a larger scale
when differential motions of galaxies relative to the nodes in the gaseous cosmic web  strip galaxies off their gas, as suggested
by Benitez-Llambay et al 2012.   As in the case of local clusters, the galactic contrails should be able to release  metal-enriched gas, perhaps enhanced by local stellar
feedback, more easily than
normal galaxies. At the very least these objects should provide a contribution to the intergalactic metal budget of galactic halos. The loss of enriched gas from galactic contrails
may suggest a solution to the long-standing puzzle of how the intergalactic medium at large distances from bright galaxies
was polluted with metals. Star formation  in galactic contrails would involve young stars, surrounded by lower HI gas columns than stars born in ordinary galaxies, and
capable of clearing their environment of dense gas, suggesting a way in which galaxies can ionize the intergalactic medium.

\section*{Acknowledgments}

The data were obtained as part of a long term collaboration with the late Wal Sargent, to whose memory we dedicate this paper.
We acknowledge helpful discussions with  Guillermo Blanc, Bob Carswell, Michele Fumagalli and Andy McWilliam.  We  thank the staff of the Keck Observatory  for their help with the observations.
MR is grateful to the National Science
Foundation for grant AST-1108815.  GB has been supported by the Kavli Foundation, and  MGH received support by the European
Research Council under the European Union's Seventh Framework Programme (FP/2007-2013) / ERC Grant
Agreement n. 320596. 
JRG acknowledges  a Millikan Fellowship at Caltech. We acknowledge use of the Atomic Line List v2.05, maintained by Peter van Hoof, 
and the use of the NIST Atomic Spectra Database (ver. 5.0; Kramida et al 2012). This research has further made use of the NASA/IPAC Extragalactic Database (NED) which is operated by the Jet Propulsion Laboratory, California Institute of Technology, under contract with the National Aeronautics and Space Administration, and of the VizieR catalogue access tool.







\bsp

\label{lastpage}

\end{document}